# Synergistical Electroluminescent Manipulation for Efficient Blue Perovskite Light-Emitting Diodes Through Interfacial Nucleation Seeding

*Hai-Yan Wu,[1,†] Yang Shen,[1,†] Yan-Qing Li,[1,2,]\* Kong-Chao Shen,[1] and Jian-Xin Tang[1,3,]\**

[1] Institute of Functional Nano & Soft Materials (FUNSOM), Jiangsu Key Laboratory for Carbon-Based Functional Materials & Devices, Soochow University, Suzhou 215123, China

[2] School of Physics and Electronics Science, Ministry of Education Nanophotonics & Advanced Instrument Engineering Research Center, East China Normal University, Shanghai, 200062, China

[3] Institute of Organic Optoelectronics (IOO), JITRI, Wujiang, Suzhou, China

\* To whom correspondence should be addressed. Email: yqli@phy.ecnu.edu.cn (Y.Q. Li), jxtang@suda.edu.cn (J.X. Tang)

[†] H.Y. Wu and Y. Shen contributed equally to this work.

**Keywords:** Perovskite light-emitting diodes; blue emission; composition engineering; defect passivation; light extraction

**Abstract**

The low efficiency of metal halide perovskite light-emitting diodes (PeLEDs) with blue emission block their potential applications in large-area full-color displays and solid-state lighting. A delicate control over the entire electroluminescence process is indispensable to overcome the efficiency limitations of blue PeLEDs. Here, we demonstrate an efficient device architecture to synergistically reduce the energetic losses during electron-photon conversion and boost the extraction of trapped light in the device. An interfacial nucleation seeding scheme is proposed to control the crystallization process of highly emissive perovskite nanocrystals and suppress the trap-mediated non-radiative recombination losses due to interfacial hydrogen bonding interactions. This manipulation results in a record external quantum efficiency (EQE) of 12.8% for blue PeLEDs emitting at 486 nm, along with the improved spectral stability and operation lifetime. Additionally, the maximum EQE reaches 16.8% after combining an internal outcoupling structure without spectral distortion, which can be further raised to 27.5% when using a lens-based structure on top of the device. We anticipate that our work provides an effective method for its application in high-performance PeLEDs.

Solution-processable metal halide perovskites hold great promise for cost-effective and high-performance light-emitting diodes due to their superior optoelectronic properties such as high photoluminescence (PL) quantum yields, widely tunable emission wavelength from blue to near infrared, high color purity, and excellent charge transport properties[1-7]. In the past few years, the device performance of perovskite light-emitting diodes (PeLEDs) has been substantially improved by manipulating the perovskite microstructures through various strategies, such as stoichiometry control, defect passivation and dimensional engineering[8-14]. Upon the optimization of the perovskite emitter, the external quantum efficiencies (EQE) over 20% has been achieved for PeLEDs with green, red, and near-infrared emissions[8-11]. Unfortunately, the low efficiency of blue PeLEDs with EQE lower than 10% is still a critical drawback, retarding their potential applications in full-color display and solid-state lighting[15-22]. It is, therefore, an urgent demand to overcome the efficiency limitations of blue PeLEDs.

To realize blue PeLEDs with high EQE, two key issues must be addressed: the perovskite emitter with blue emission should be as efficient as possible (high internal quantum efficiency), and a large portion of internally created photons must be extracted into air (high outcoupling efficiency). Previously, the research community has focused on optimizing the perovskite emitters through the perovskite precursor engineering, including the compositional modulation to tune the emission colors, the dimensional control with additives to enhance the radiative recombination efficiency, and the defect passivation to suppress the non-radiative recombination loss[20-24]. For example, the quasi-two-dimensional (quasi-2D) Ruddlesden–Popper structures can utilize the quantum confinement to promote the internal energy funnel inside the perovskite film, leading to the improved electroluminescence (EL) efficiency for blue PeLEDs[20-22]. However, recent works on perovskite optoelectronic devices have unveiled the significant impacts of the underlying interlayers on the perovskite crystallization process, which is critical to the film quality and device performance[25-28]. To unlock the full potential of blue PeLEDs, tailoring the surface chemical and physical properties of the buried interlayers may provide an effective way for boosting the efficiency of blue PeLEDs. In addition, a stratified thin-film structure of PeLEDs suffer from the inferior outcoupling efficiency due to index-mismatch-induced total internal reflection (TIR). The outcoupling efficiency of the conventional PeLEDs with a planar structure is usually limited to only 20-30%[29-32]. Therefore, a substantial increase in EQE is expected if the optical management can be taken into account to enhance the outcoupling of the trapped light in PeLEDs.

Here we propose an efficient device architecture for blue PeLEDs that combines a well-controlled formation of highly emissive perovskite layer with an improved outcoupling structure. An interfacial nucleation seeding (INS) scheme is implemented for driving the crystallization process of inorganic perovskite emitter on top of a carefully modulated hole transport layer (HTL), where interfacial hydrogen bonding interactions with amino and hydroxyl groups are introduced to provide the nucleation sites and retard the grain sizes of the perovskite emitter. As a result, small crystalline grains with reduced trap states are formed in perovskite films, leading to the spatial limit for exciton diffusion and the suppression of non-radiative recombination loss. A peak EQE of 12.8% is thus obtained for blue PeLEDs emitting at 486 nm, along with the improved spectral stability and operation lifetime. Apart from the reduced energetic loss during electron-photon conversion, the extraction of the waveguided light can be significantly boosted by patterning an internal outcoupling nanostructure on the buried HTL, resulting in the maximum EQE of 16.8% without spectral distortion. Additionally, the EQE can be further raised to 27.5% when using a lens-based outcoupling structure on top of the device to reduce the amount of light trapped in the substrate mode.

**Results**

**Composition Engineering for Blue Emission.** To obtain the blue emission, the composition engineering was used by controlling the Br:Cl ratios in all inorganic $CsPbBr_{3-x}Cl_x$ perovskite films to enlarge their emission bandgap[21,33]. As illustrated in Supplementary Fig. S1, $CsPbBr_{3-x}Cl_x$ perovskites were processed through spin-coating from stoichiometrically modified precursors with different molar ratios of lead bromide ($PbBr_2$) to lead chloride ($PbCl_2$) ($PbBr_2:PbCl_2$ = 1.5:1.0, 1.0:1.0, and 1.0:1.5). The total molar ratio of CsBr to ($PbBr_2$ + $PbCl_2$) was fixed to 1.3:1.0 in the precursor solution, where the CsBr excess was used to eliminate the non-perovskite δ-phase. To passivate the defects and confine the quasi-2D structures in perovskite films, the additives of polyethyleneglycol (PEG), potassium bromide (KBr), phenylethylammonium bromide (PEABr), and 4-Fluorophenylethylammonium bromide (p-F-PEABr) were added in the precursor solutions (see the details in Methods)[34,35].

As characterized by scanning electron microscope (SEM) (Supplementary Fig. S2), $CsPbBr_{3-x}Cl_x$ perovskite films with different $PbBr_2$:$PbCl_2$ ratios exhibit the similar morphologies with numerous pinholes and defects at the grain boundaries. The grain sizes of different $CsPbBr_{3-x}Cl_x$ perovskite films are slightly affected by the varied $PbBr_2$:$PbCl_2$ ratios in the perovskite precursor solutions. Meanwhile, the X-ray diffraction (XRD) patterns (Supplementary Fig. S3) reveal that the crystal structures of $CsPbBr_{3-x}Cl_x$ perovskite films processed with different $PbBr_2$:$PbCl_2$ ratios are dominated by orthorhombic phase, which are in accordance with previous report[36]. The chemical states and elemental stoichiometries in $CsPbBr_{3-x}Cl_x$ perovskite films processed with different $PbBr_2$:$PbCl_2$ ratios were analyzed by X-ray photoelectron spectroscopy. It is confirmed that all the elements (*e.g.*, Cs, Pb, Br, and Cl) show the single chemical state (Supplementary Fig. S4), and a gradual decrease in Br:Cl atomic ratio is observed when tuning the $PbBr_2$:$PbCl_2$ molar ratio from 1.5:1.0 to 1.0:1.5 (Supplementary Table S1). However, the (Br + Cl):Pb atomic ratio in $CsPbBr_{3-x}Cl_x$ perovskite films also decreases when increasing the $PbCl_2$ molar proportion in the precursors. This change is associated with the relatively poor solubility of $PbCl_2$ in dimethyl sulfoxide (DMSO) that limits the Cl concentration in the perovskite films.

The optical properties of $CsPbBr_{3-x}Cl_x$ perovskite films processed with different $PbBr_2$:$PbCl_2$ ratios were characterized by steady-state PL measurements. As compared in Figure 1a, an increase in $PbCl_2$ molar proportion in perovskite precursors gives rise to a gradual hypsochromic shift of emission colors from green to blue for $CsPbBr_{3-x}Cl_x$ perovskite films. The PL peaks are centered at 491, 484 and 477 nm for the $PbBr_2$:$PbCl_2$

molar ratios of 1.5:1.0, 1.0:1.0 and 1.0:1.5, respectively. When these perovskite films were used to fabricated the PeLEDs (see **Methods** for details), the corresponding EL spectra are almost identical to their PL spectra (Figure 1a). It is evident that blue PeLEDs exhibit the almost symmetric peaks, which were centered at 494, 486 and 478 nm, respectively, with a narrow full width at half maximum of 20 nm. As illustrated by the Commission Internationale de l'Eclairage (CIE) color coordinates in Figure 1b, PeLEDs using $CsPbBr_{3-x}Cl_x$ perovskite films can emit in the blue region with high color purity. While the emission colors can be precisely modulated through composition engineering, the dramatically degraded device performance is accompanied with the increased Cl molar proportion in $CsPbBr_{3-x}Cl_x$ perovskite films (Supplementary Fig. S5). For instance, the EQE values of PeLEDs are 7.7%, 5.4%, and 3.3% when using $CsPbBr_{3-x}Cl_x$ perovskite films with $PbBr_2:PbCl_2$ molar ratios of 1.5:1.0, 1.0:1.0 and 1.0:1.5, respectively (Supplementary Table S2).

**Crystallization Engineering for Emissive Perovskites.** It has been identified that one of the major efficiency limits for PeLEDs is the severe trap-mediated non-radiative recombination losses at the contact interfaces and around the grain boundaries of perovskite films due to ionic defects (like halide vacancies)[4,37,38]. Moreover, it has been demonstrated that small crystalline grains are needed for emissive perovskite films and efficient PeLEDs, which can spatially limit the exciton diffusion length and maximize the possibility of exciton radiative recombination[5,6,36]. It is well known in the film growth theory that the initial seeds play a dominant role in the crystallization kinetics of the final films.

To improve the perovskite film quality and device performance, the crystallization engineering was carried out by introducing a new INS method. Figure 2 illustrates the proposed mechanism of an INS-triggered crystallization process of perovskite films on the underlying substrates with different chemical and physical properties. The crystallization of perovskite films commonly takes place from the nucleation via the chemical reaction among CsBr, $PbBr_2$ and $PbCl_2$ in the precursor solution on the substrate, followed by the annealing-assisted grain growth through interdiffusion and complete reaction of different precursors. For non-wetting substrates (Figure 2a), the growth of large perovskite grains is expected due to the presence of high grain boundary mobility. On the contrary, the addition of some molecules with special functional groups (*e.g.*, $-NH_2$ or -OH) can modulate the substrate surface with better wettability[25,26,39-41]. The coating of perovskite precursors with high surface coverage will be promoted due to the large surface tension dragging force on the wetting surface (Figure 2b). More importantly, the hydrogen bonding interactions may take place between the functional groups (*e.g.*, $-NH_2$ and -OH) and halide ions[27,39,42,43], which can provide the nucleation sites and suppress the crystal growth with reduced trap states. Consequently, the growth of perovskite grains will be spatially limited due to the dense nuclei from heterogeneous nucleation and the pinned grain boundaries with the excess cations or additives, and a well-packed assembly of small grains is expected be achievable for compact and pinhole-free perovskite films on the wetting substrate.

To verify the effectiveness of the proposed INS method for the crystallization engineering of perovskite films, the commonly used HTL of poly(3,4-ethylenedioxythiophene):poly(styrenesulfonate) (PEDOT:PSS) was modified by adding an optimized amount (0.2 vol%) of amino-functionalized ethanolamine (ETA) (see **Methods** for the fabrication details). The impacts of ETA-modified PEDOT:PSS (denoted as EMP) on film quality and device performance were systematically characterized for $CsPbBr_{3-x}Cl_x$ perovskite films with a $PbBr_2$:$PbCl_2$ molar ratio of 1.0:1.0. The measured contact angles display a decrease from 34.2° to 18.5° when a conventional PEDOT:PSS HTL is modified with an ETA additive (Figure 3a), indicating the more hydrophilic feature of the EMP surface and the enhanced wetting capability for the perovskite precursor solution. The improved wettability may be ascribed to the presence of ETA-introduced amino and hydroxyl groups on the PEDOT:PSS surface. SEM images in Figure 3b,c show a clear transition for film morphologies of $CsPbBr_{3-x}Cl_x$ perovskites grown on two HTLs, revealing smoother surface with reduced pinholes and smaller grain sizes on the EMP substrate. The average grain size is reduced from 26.5 nm to 20.2 nm for the $CsPbBr_{3-x}Cl_x$ perovskite grown on the EMP substrate (Supplementary Fig. S6). In addition, the enhanced crystallinity is observed when depositing the perovskite film on EMP as indicated by the synchrotron-based 2D grazing incidence XRD (GIXRD) images (Figure 3d,e). The obvious increase in crystal (200) and (210) diffractions confirms the influence of INS-triggered crystallization for perovskite nanocrystals.

The luminescent properties of $CsPbBr_{3-x}Cl_x$ perovskites grown on different substrates were characterized by steady-state and transient PL measurements. It is evident that CsPbBr$_{3-x}$Cl$_x$ perovskite film grown on EMP exhibits the higher PL intensity than that on conventional PEDOT:PSS (Supplementary Fig. S7). Meanwhile, the CsPbBr$_{3-x}$Cl$_x$ perovskite film grown on EMP shows the prolonged PL decay curves (Figure 3f), indicating the suppression of non-radiative recombination losses. The enhanced luminescent properties are coincident with the observation of the INS-triggered growth of highly crystalline perovskite films.

To evaluate the effect of INS-triggered perovskite crystallization on EL performance, blue PeLEDs were fabricated as schematically illustrated in Figure 4a. The device architecture is composed of glass/indium tin oxide (ITO)/HTL (40 nm)/perovskite (35 nm)/2,2',2"-(1,3,5-Benzinetriyl)-tris(1-phenyl-1-H-benzimidazole) (TPBi) (50 nm)/lithium fluoride (LiF) (1 nm)/aluminum (Al) (100 nm). Here, conventional PEDOT:PSS and EMP were used as a HTL for comparison, while TPBi was used as electron-transport layer (ETL). ITO was used as an anode and LiF/Al was used as a bilayer cathode. Device characteristics of blue PeLEDs are plotted in Figure 4b-4f, and key performance parameters are summarized in Table 1. Current density-voltage (J-V) and luminance-voltage (L-V) curves indicate that blue PeLED with an INS-triggered perovskite emitter exhibits both improved electrical and luminance properties as compared to a control device (Figure 4b). It is noted that the turn-on voltage (V$_{ON}$, that is defined at a luminance of 1 cd m$^{-2}$) of PeLEDs is largely decreased from 4.2 V of the control device to 3.6 V of the INS-triggered one, indicating the improved EMP/perovskite contact for enhancing hole injection. The identical EL spectra are observed for both devices with a narrow emission peak centered at 486 nm (Figure 4c), and the INS-triggered device presents the spectral stability under various bias voltages (Supplementary Fig. S8). However, it is evident in Figure 4d that the use of an INS-triggered perovskite film causes a substantial increase in device efficiency. For example, the maximum EQE and current efficiency (CE) of blue PeLED with an INS-triggered perovskite film are significantly boosted to 12.8 % and 15.4 cd A$^{-1}$ (Table 1), respectively, which are about two times higher than that of the control device fabricated on a conventional PEDOT:PSS (*e.g.*, EQE = 6.3% and CE = 7.8 cd A$^{-1}$). These efficiencies represent the highest values ever reported for blue PeLEDs (Supplementary Table S3)[15,18,21,22,34,44-47]. In addition, the histograms of peak EQEs displayed in Figure 4e reveals high reproducibility of blue PeLEDs, and the use of INS-triggered perovskite films results in an average EQE of 10.3%. It is known that the EL process is domniated by carrier and exciton dynamics. According to SEM, GIXRD and PL results in Figure 3, it is thus inferred that such remarkable enhancements of EQE and CE are correlated to the suppression of both trap-mediated non-radiative recombination loss and exciton quenching due to the INS-triggered growth of highly crystalline perovskite films with reduces grain sizes for stronger spatial confinement of exciton radiative recombination. Moreover, the operational stability of blue PeLEDs were studied under a constant current density. It is noteworthy that the use of an INS-triggered perovskite film induces the significant improvement in the operating stability (Figure 4f). The half-lifetime at an initial luminance of 150 cd m$^{-2}$ is increased from 12.6 min of the control device to 16.2 min of INS-triggered one.

**Optical Engineering for Light Extraction.** Apart from the optimized perovskite emitter for efficient radiative recombination during electron-photon conversion, the optical engineering was further implemented to enhance the outcoupling efficiency of blue PeLEDs. As indicated by the theoretical simulation (Supplementary Fig. S9), a large fraction of the emitted light from the perovskite emitter will be trapped in the waveguide mode due to an index-mismatched sandwich structure of high-refractive-index perovskite ($n_{pero}$ = 2.3)/low-refractive-index PEDOT:PSS HTL ($n_{HTL}$ = 1.5)/high-refractive-index ITO ($n_{ITO}$ = 2.0). On the contrary, the sub-wavelength nanopatterning at the HTL/perovskite interface causes the remarkable outcoupling enhancement for the emitted light from the waveguide mode into the substrate mode, and partial of the waveguided light in a planar device architecture can effectively enter the glass substrate (Supplementary Fig. S9). Inspired by the simulation results, aperiodic funnel-shaped nanoarrays were integrated into the EMP HTL by adopting the soft nanoimprint technique with a poly(dimethylsiloxane) (PDMS) mold for the internal light extraction (iLE) from the perovskite emitter into the glass substrate (see the details in **Methods**)[48]. Figure 5a and 5b display the cross-sectional SEM images of planar and patterned devices with the iLE nanostructures. As shown in Figure 5b and Figure S10 of Supporting Information, the iLE nanostructures were transferred to the EMP HTL, featuring an average period of ~300 nm and groove depth of ~25 nm. The $CsPbBr_{3-x}Cl_x$ perovskite film can form a compact contact with the patterned EMP HTL, and the corrugated structures are well preserved at the patterned HTL/perovskite interface (Figure 5b). However, the top surface of $CsPbBr_{3-x}Cl_x$ perovskite film exhibits a smooth pinhole-free morphology, which is almost identical to that formed on a planar HTL (Supplementary Fig. S10).

Figure 5c plots the J-V and L-V curves of the patterned device with an iLE structure. Compared to the planar devices shown in Figure 4a, the patterned device shows a reduced $V_{ON}$ of 3.3 V and a higher maximum luminance of 1694 cd m$^{-2}$ (Table 1). In addition, the EL spectrum of the patterned device shows an identical emission centered at 486 nm, and the angular dependence of normalized EL intensities follows an ideal Lambertian profile (Figure 5d). Remarkably, the peak EQE and CE of the patterned device with an iLE structure are boosted to 16.8% and 20.4 cd A$^{-1}$, respectively (Figure 5e). The EQE histogram in Figure 5f indicates a high reproducibility with an average value of 14.7%. These results illustrate a great potential of iLE nanostructures for outcoupling the waveguided light in blue PeLEDs. As an amount of the light entering the glass substrate is trapped in substrate mode due to TIR at the glass/air interface, the external light extraction (eLE) structures are required for further enhancing the light output. An index-matched glass half-ball lens was used as an eLE structure, which enables the outcoupling of light under high incidence angles[49]. It is apparent in Figure 5 that the synergetic interplay of iLE and eLE structures results in a substantial increase in EQE and CE to 27.5% and 33.4 cd A$^{-1}$ (Figure 5e), respectively, along with an average EQE of 26.3% (Figure 5f). To our best knowledge, these values represent the record performance for PeLEDs with blue emission (Supplementary Table S3).

**Discussions**

Our results demonstrate that highly efficient blue PeLEDs are achievable by combining the strategies of crystallization engineering and light manipulation for a delicate control over the entire electroluminescence process. An interfacial nucleation seeding strategy is implemented for driving the crystallization process of highly emissive perovskite emitters, resulting in the formation of small crystalline grains with reduced trap states via interfacial hydrogen bonding interactions. The energetic losses during electron-photon conversion are significantly reduced due to the spatial limit for exciton diffusion and the suppressed non-radiative recombination in the perovskite film. A peak EQE of 12.8% is achieved for blue PeLEDs emitting at 486 nm along with the improved spectral stability and operation lifetime. Moreover, the extraction of trapped light in the device is substantially boosted due to a synergetic interplay of internal and external extraction structures. A maximum EQE of 27.5% is thus obtained by largely releasing the amount of light trapped in both waveguide and substrate modes. We anticipate that our method paves the way for the realization of high-performance PeLEDs as novel light sources in full-color displays and solid-state lighting applications.

**Methods**

**Materials and Chemicals.** Cesium bromide (CsBr, 99.0%) and lead bromide (PbBr$_2$, 99.0%) were purchased from TCI. Phenylethylammonium bromide (PEABr, 99.5%) was purchased from Borun Corp. Potassium bromide (KBr, 99.5%) and ethanolamine (ETA, 99.0%) were purchased from Aladdin Industrial Corporation. Lead chloride (PbCl$_2$, 99.99%) and 4-Fluorophenylethylammonium bromide (p-F-PEABr, 99.5%) were purchased from Xi'an Polymer Light Technology Corp. Polyethyleneglycol (PEG, average Mv ≈ 600 000) and lithium fluoride (LiF, 99.99%) were purchased from MACKLIN. 2,2′,2″-(1,3,5-benzinetriyl)-tris(1-phenyl-1-H-benzimidazole) (TPBi, 99.0%) was purchased from Nichem. Dimethyl sulfoxide (DMSO, 99.9%) was purchased from Alfa Aesar. All the above chemicals were directly used without any further purification.

**Perovskite Precursor Solutions.** The perovskite precursor solutions were prepared by dissolving CsBr (0.2 mmol mL$^{-1}$), PbBr$_2$ and PbCl$_2$ with a molar ratio of 2.6:1:1, together with p-F-PEABr (6.67 wt%), PEABr (3.33 wt%) and PEG (2.31 wt%) in DMSO. For the precursor with KBr passivation, KBr at a concentration of 2 mg mL$^{-1}$ was added into the solution. All the prepared precursor solutions were stirred at 40 °C for a few hours and filtered with 0.45 μm poly(tetrafluoroethylene) syringe filters.

**Substrate Preparation.** Indium-tin-oxide (ITO)-coated glass substrates were cleaned by ultrasonication successively in detergent, deionized water, acetone, ethanol, isopropyl alcohol (each step for 5 min), and then dried in an oven. Prior to film deposition, ITO-coated glass substrates were further treated with UV ozone for 20 min to enhance the surface hydrophilicity. The filtered poly(3,4-ethylenedioxythiophene):poly(styrenesulfonate) (PEDOT:PSS, Clevios Al 4083) layer was spin-coated on ITO glass substrate at 4000 rpm, and subsequently annealed at 140 °C in ambient air. For the nucleation seeding, the PEDOT:PSS solution was modified by doping the ETA molecules with a concentration varying from 0.1 vol% to 0.4 vol%.

**Outcoupling Structures.** The poly(dimethylsiloxane) (PDMS) mold of aperiodic funnel-shaped nanoarrays with an average period around 300 nm was prepared as reported previously[48]. After spin-coating the PEDOT:PSS (or ETA-modified PEDOT:PSS) on ITO glass substrate and annealing the sample at 140 °C for 60 s in ambient air, the PDMS mold was pressed onto the PEDOT:PSS layer under a constant pressure of 0.5 bar for 30 s. Then, the mold was peeled off, and PEDOT:PSS film was further annealed for 15 min.

**Device Fabrication.** The perovskite films were formed in a nitrogen-filled glovebox by spin-coating the precursor solutions at 3000 rpm for 60 s onto PEDOT:PSS, and immediately annealed at 70 °C for 8 min. Then, the samples were transferred into an interconnected high-vacuum deposition system (Trovato Co., base pressure $< 3 \times 10^{-7}$ Torr), in which TPBi (50 nm), LiF (1 nm) and Al (100 nm) were successively deposited via thermal evaporation with shadow masks. The deposition rate and film thickness were monitored with a quartz crystal oscillator. The active area of PeLEDs was determined to be 10 mm$^2$ from the overlap between bottom ITO and top Al electrodes. To ensure the consistent experiment conditions, planar and patterned PeLEDs were fabricated in the same batch.

**Film and Device Characterizations.** Surface morphologies of various films were characterized by scanning electron microscopy (SEM) (Zeiss Supra 55). X-ray diffraction (XRD) patterns were recorded by X-ray diffractometer (PANalytical X'Pert

Pro) with monochromatic Cu Kα radiation (λ = 0.154 nm). The synchrotron-based grazing incidence XRD (GIXRD) measurements were performed at the BL14B1 beamline of the Shanghai Synchrotron Radiation Facility using X-ray with a wavelength of 1.24 Å and a grazing incidence angle of 0.2°. X-ray photoelectron spectroscopy (XPS) measurements were performed with a Kratos AXIS Ultra-DLD ultrahigh vacuum photoemission spectroscopy system using a monochromatic Al Kα source (1486.6 eV) with a total instrumental energy resolution of 500 meV. Steady-state photoluminescence (PL) spectra were measured with a FluoroMax-4 fluorescence spectrometer (Horiba Jobin Yvon) under the ambient environment. Transient PL decay measurements were performed with a Quantaurus-Tau fluorescence lifetime spectrometer (C11367-32, Hamamatsu Photonics) under vacuum atmosphere with a 373 nm pulsed laser (pulse width of 100 ps and repetition rate of 5 KHz). Layer thickness, frequency-dependent refractive index ($n$), and extinction coefficient ($k$) values of different films were measured by an alpha-SE™ Spectroscopic Ellipsometry with an angle of incidence at 70° (J. A. Woollam Corp). Contact angles were measured with a contact angle tester (DataPhysics instruments GmbH). Current density-voltage-luminance (J-V-L) characteristics and electroluminescence (EL) spectra of PeLEDs were measured simultaneously with a computer-controlled programmable power source (Keithley model 2400) and a luminance meter/spectrometer (PhotoResearch PR670) in air at room temperature. Angular EL measurements were conducted by rotating the sample stage. The EQE values were obtained by the measurements of luminous flux with an integrating sphere (Hamamatsu Photonics K.K. C9920-12). The operational lifetimes of PeLEDs were measured with a 64-channel ZJLS-4 type OLED-aging testing system in constant-current mode.

**Acknowledgements**

This work is financially supported by the National Natural Science Foundation of China (Nos. 61520106012, 61722404, 51873138, 91633301), the National Key R&D Program of China (Nos. 2016YFB0401002, 2016YFB0400700), the 333 Program (No. BRA2019061), and the Collaborative Innovation Center of Suzhou Nano Science & Technology.

**Author contributions**

Y.L. and J.T. conceived the idea and designed the experiments. H.W. performed the device fabrication and property characterization. Y.S. performed the nanopatterns, optical characterization and simulations. K.S. contributed to the measurements of crystal structures. Y.L. analyzed most of the data and wrote the manuscript. J.T. motivated this work and co-wrote the manuscript. All authors were involved in extensive discussions and data analyses.

**Additional information**

Correspondence and requests for materials should be addressed to Y.L. and J.T.

**Competing financial interests**

The authors declare no competing financial interests.

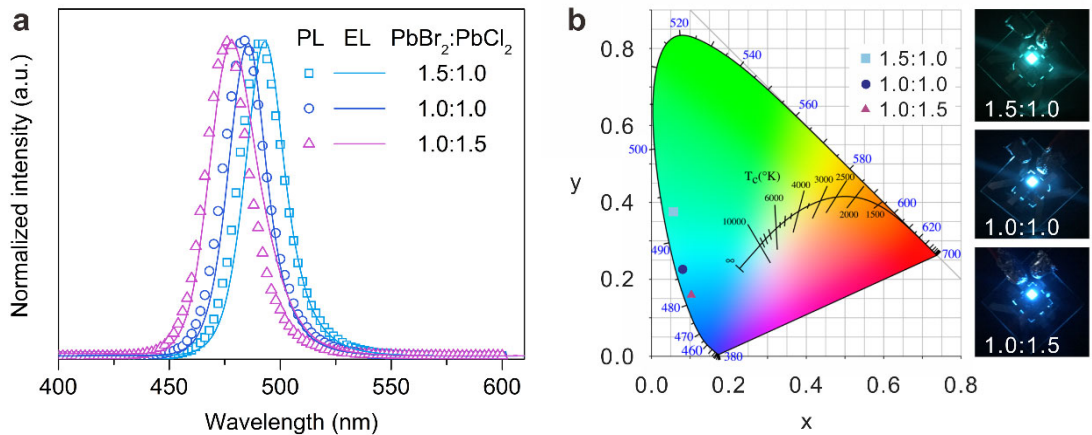

**Figure 1.** Composition engineering for CsPbBr$_{3-x}$Cl$_x$ perovskite films processed from precursors with different molar ratios of lead bromide (PbBr$_2$) to lead chloride (PbCl$_2$). a) Normalized PL spectra (open symbols) of perovskite films and EL spectra (solid lines) of PeLEDs. b) Commission Internationale de l'Eclairage (CIE) coordinates of the EL spectra of PeLEDs with the photos of the operating devices.

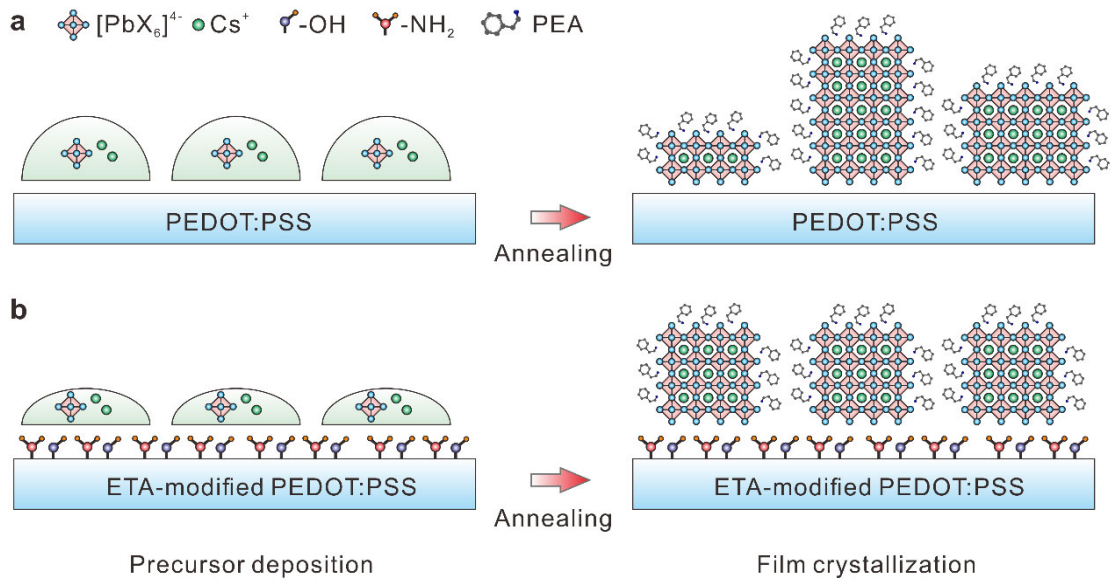

**Figure 2.** Schematic illustration of the INS-triggered perovskite crystallization. The nucleation and growth of $CsPbBr_{3-x}Cl_x$ perovskite films on (a) conventional and (b) ETA-modified PEDOT:PSS HTLs, respectively.

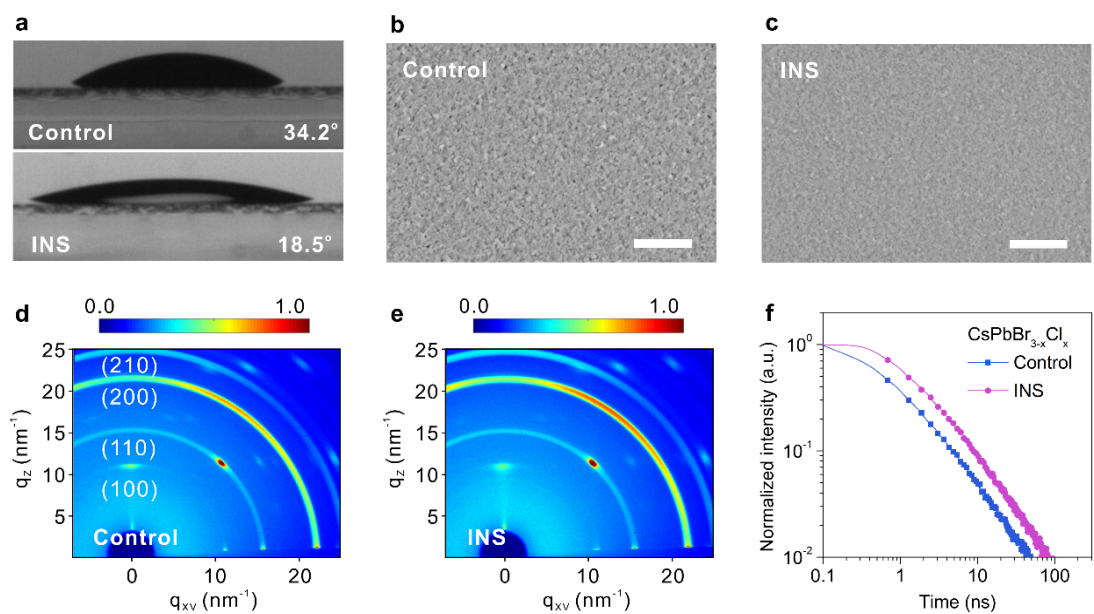

**Figure 3.** INS-triggered crystallization engineering for CsPbBr$_{3-x}$Cl$_x$ perovskite films. a) Contact angles of precursor solutions on PEDOT:PSS and EMP. b,c) Top-view SEM images of CsPbBr$_{3-x}$Cl$_x$ perovskite films grown on (b) PEDOT:PSS and (c) EMP. Scale bar in SEM images is 200 nm. d,e) GIXRD images of CsPbBr$_{3-x}$Cl$_x$ perovskite films grown on (d) PEDOT:PSS and (e) EMP. f) Transient PL decay curves of different CsPbBr$_{3-x}$Cl$_x$ perovskite films.

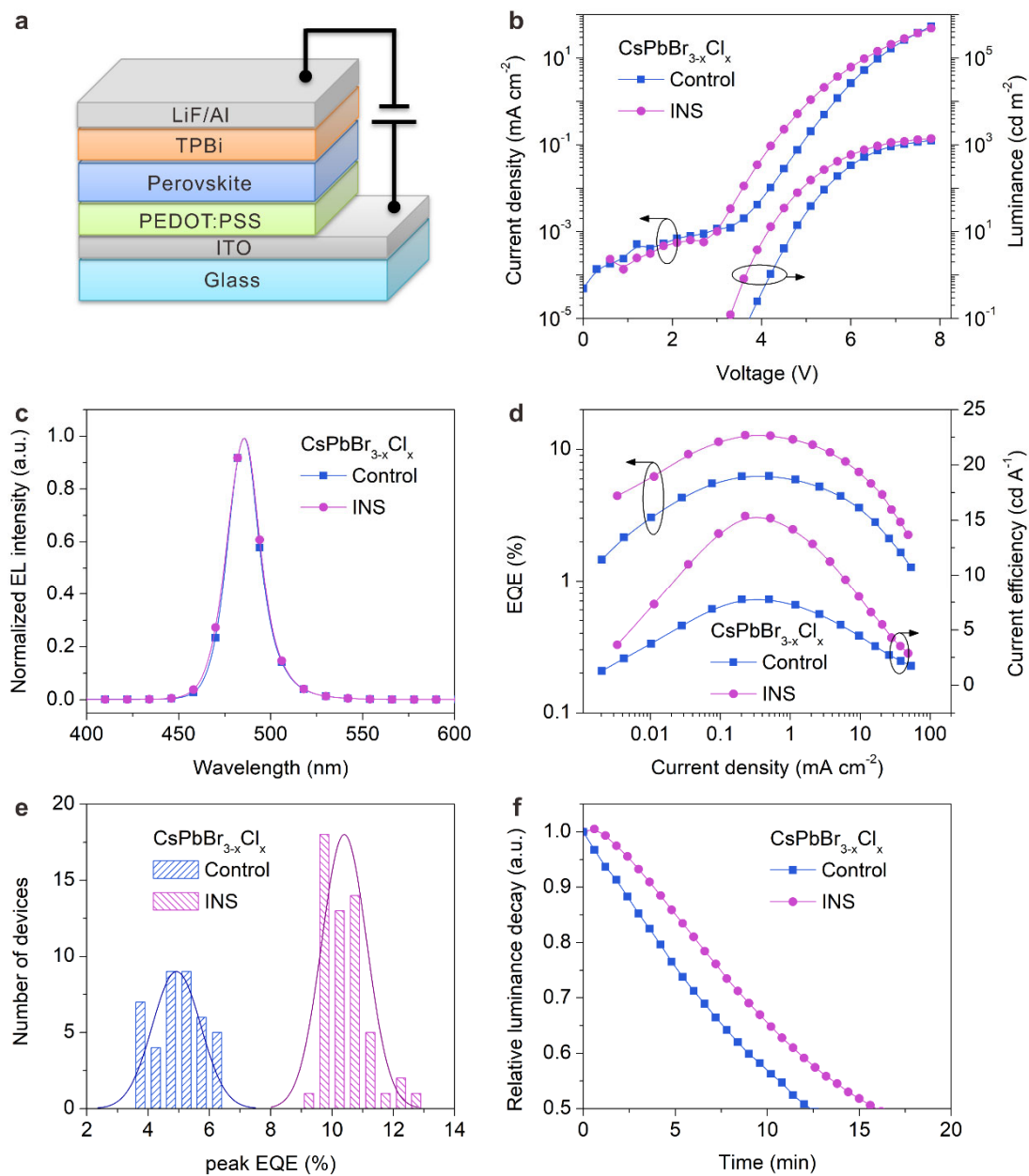

**Figure 4.** Device structure and performance characteristics of blue PeLEDs with crystallization engineering. a) Schematic diagram of the device architecture. b) Current density-voltage (J-V) and luminance-voltage (L-V) curves. c) Normalized EL spectra. d) EQE and current efficiency as a function of current density. e) Histograms of peak EQEs of different device structures. f) Operating lifetimes under constant current density with an initial luminance of 150 cd m$^{-2}$.

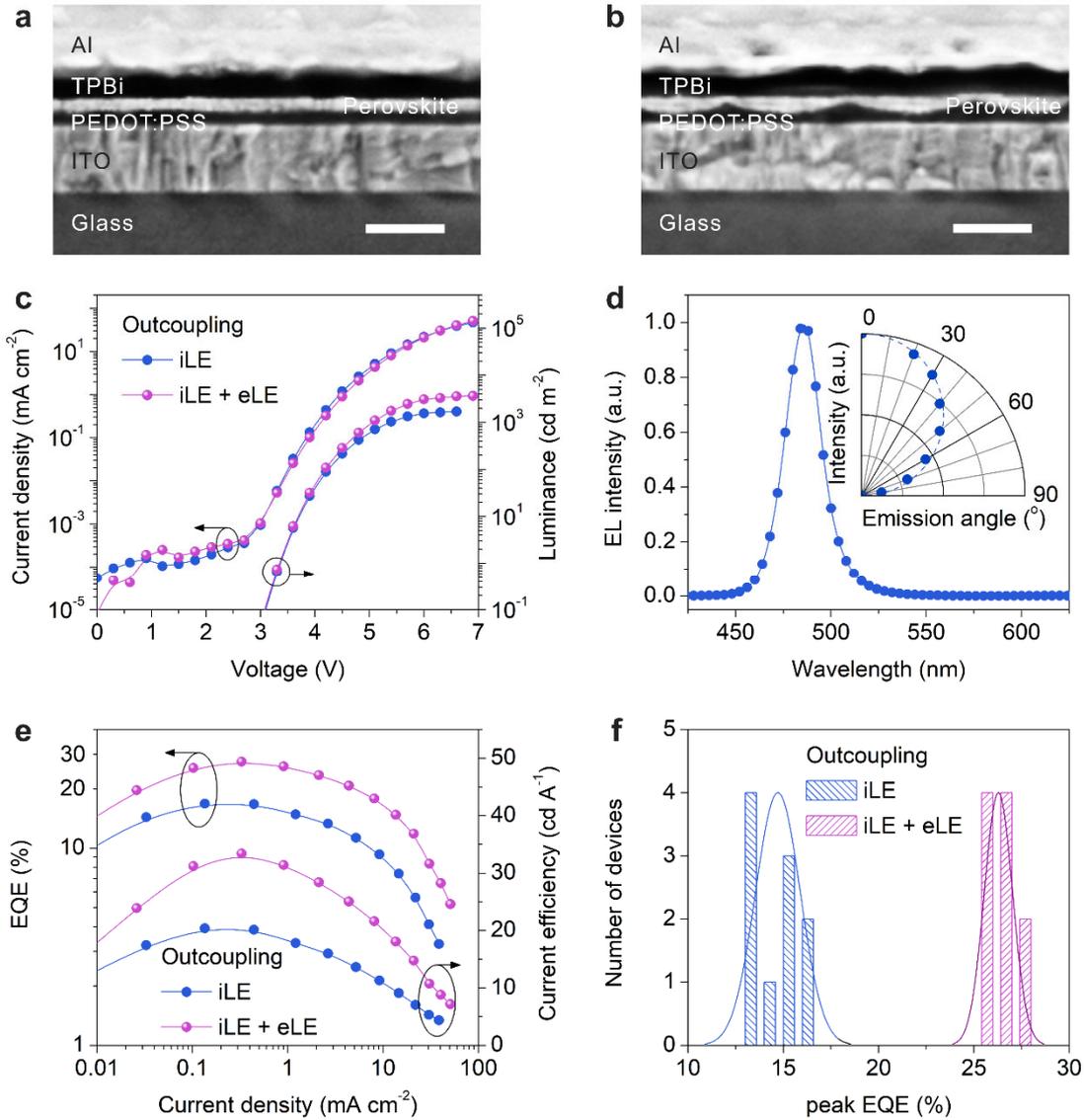

**Figure 5.** Light manipulation and device performance of blue PeLEDs. a,b) Cross-sectional SEM images of (a) planar and (b) patterned devices with internal light extraction (iLE) nanostructures (Scale bar: 200 nm). c) J-V and L-V curves of patterned devices with iLE and external light extraction (eLE) lens. d) EL spectrum of the patterned device with iLE. Inset is the polar plot of normalized angular EL intensities (solid symbols) compared to an ideal Lambertian pattern (dashed line). e) EQE and CE a function of current density. f) Histogram of peak EQEs.

**Table 1.** Device performance of CsPbBr$_{3-x}$Cl$_x$-based blue PeLEDs with different modifications.

| Device structures | V$_{ON}$ [V] | EL peak [nm] | Max L [cd m$^{-2}$] | EQE [%] | CE [cd A$^{-1}$] | CIE [x, y] |
|---|---|---|---|---|---|---|
| Control | 4.2 | 486 | 1243 | 6.3 | 7.8 | [0.081, 0.224] |
| INS | 3.6 | 486 | 1390 | 12.8 | 15.4 | [0.082, 0.219] |
| INS+iLE | 3.3 | 486 | 1694 | 16.8 | 20.4 | [0.083, 0.221] |
| INS+iLE+eLE | 3.3 | 486 | 3651 | 27.5 | 33.4 | [0.083, 0.224] |